\documentclass[12pt]{article}

\usepackage{amsmath}
\usepackage{amsfonts,amssymb,amsthm,overpic}
\makeatletter
\newcommand{\diam}{\mathop{\operator@font diam}}
\usepackage{setspace}
\usepackage{multicol}
\usepackage{multirow}
\usepackage[compact]{titlesec}

\usepackage{epsfig}

\usepackage{xcolor}

\newtheorem{definition}{Definition}[section]

\newtheorem{lemma}{Lemma}[section]

\newcommand{\cT}{\mathcal{T}}

\begin{document}

\title{\Huge{\textsc{On the Causal and Topological Structure of the $2$-Dimensional Minkowski Space.}}}

\author{Kyriakos Papadopoulos$^1$, Nazli Kurt$^2$, Basil K. Papadopoulos$^3$\\
\small{1. Department of Mathematics, Kuwait University, PO Box 5969, Safat 13060, Kuwait}\\
\small{2. Open University, UK}\\
\small{3. Department of Civil Engineering, Democritus University of Thrace, Greece}\\
E-mail: \textrm{ kyriakos@sci.kuniv.edu.kw}
}

\date{}

\maketitle

\begin{abstract}
A list of all possible causal relations in the $2$-dimensional Minkowski space $M$ is exhausted,
based on the duality between timelike and spacelike in this particular case, and thirty topologies
are introduced, all of them encapsulating the causal structure of $M$. Generalisations
of these results are discussed, as well as their significance in a discussion on
spacetime singularities.
\end{abstract}

\section{Preliminaries.}

Throughout the text, unless otherwise stated,
we consider the two-dimensional Minkowski spacetime $M$, that
is the two-dimensional real Euclidean space equipped
with the characteristic quadratic form $Q$, where for
$x =(x_0,x_1)\in M $, $Q(x) = x_0^2 -x_1^2$.

We denote the {\em light cone} through an event $x$
by $C^L(x)$, and define it to be the set $C^L(x)=\{y : Q(y-x)=0\}$.
Similarly, we define the {\em time cone} as $C^T(x) = \{y: y=x \textrm{ or }
Q(y-x)>0\}$ and the {\em space cone} as $C^S(x) = \{y: y=x \textrm{ or }
Q(y-x)<0\}$. We call {\em causal cone} the set $C^T(x) \cup C^L(x)$ and we
observe that the event $x$ partitions its time/light/causal cone into
future and past time/light/causal cones, respectively, while it
divides the space cone into $-$ and $+$, respectively.

In \cite{On-Two-Zeeman-Topologies} (paragraph 1.4), we intuitively (i.e. in a 
topological sense, invariantly from a change in the geometry) partitioned
the light-cone so that apart from future and past we also achieved a spacelike
separation of $+$ and $-$. This space-like separation is more obvious
in the $2$-dimensional Minkowski spacetime $M$. Let $x \in M$ be an event.
Then, we consider the future and past time-cones, $C_+^T(x)$ and
$C_-^T(x)$, respectively, as North and South in a compass, while
the space-cones $C_+^S(x)$ and $C_-^S(x)$, respectively, as East and
West.

We denote the Euclidean topology on $\mathbb{R}^2$ by $E$; this topology
has a base of open sets which are open balls $B_\epsilon(x)$, of radius $\epsilon$ and centre
$x$. Arbitrary unions of such open balls give the open sets in $\mathbb{R}^2$ under $E$.

Zeeman, in \cite{Zeeman1} (as a result of his previous work in \cite{Zeeman2}) has questioned
the use of the topology $E$ in $4$-dimensional Minkowski space, as its ``natural'' topology,
listing a number of issues, including that the Euclidean topology is locally homogeneous
(while $M$ is not) and the group of all homeomorphisms of (four dimensional) Euclidean
space is of no physical significance. Zeeman proposed a topology, his ``Fine'' topology,
under which the group of all homeomorphisms is generated by the (inhomogeneous) Lorentz
group and dilatations. In addition, the light, time and space cones through a point can be deduced
from this topology.
 G\"obel, in \cite{gobel}, generalised Zeeman's results for
curved spacetime manifolds, and obtained that under a general relativistic frame,
the Fine topology gives the significant result that a homeomorphism is an isometry.
Hawking-King-McCarthy, in \cite{Hawking-Topology}, introduced the ``Path'' topology,
which determines the causal, differential and conformal structure of a space-time,
but it was proven by Low, in \cite{Low_path}, that the Limit Curve Theorem under the
Path topology fails to hold, and so the formation of basic singularity theorems. Given
that the questions that were raised by Zeeman in \cite{Zeeman1} are of a tremendous
significance for problems related to the topological, geometrical and analytical
structure of a spacetime, the topologisation problem for spacetimes is still open
and significant.

In this article, we examine all possible (ten in number) causal relations which can appear in the
$2$-dimensional Minkowski spacetime and the thirty topologies which they induce. All these
topologies incorporate the causal structure of spacetime, and we believe that a generalisation
to curved $4$-dimensional spacetimes will equip modern problems of general relativity and cosmology
with extra tools, that can be used in attempts, for example, to describe the structure
of the universe in the neighbourhood of the spacetime singularities that are 
predicted by the singularity theorems of general relativity (ambient cosmology)
or contribute to the description of the transition from the quantum non-local
theory to  a classical local theory.

\section{Causal relations in the $2$-Dimensional Minkowski Space.}

We consider the $2$-Dimensional Minkowski Spacetime $M$, equipped
with the following relations:
\begin{enumerate}
\item $\ll$; the {\em chronological} partial order, defined as $x \ll y$, if $y \in C_+^T(x)$. We
note that $\ll$ is irreflexive.

\item $\rightarrow$; the relation {\em horismos}, defined as $x \rightarrow y$, if $y \in C_+^L(x)$. Horismos
is a reflexive relation.

\item $<$; the {\em chorological} (``choros'' is the Greek for ``space'', just like ``chronos'' is
the Greek for ``time'') partial order, defined as $x < y$, if $y \in C_+^S(x)$. We note
that $<$ is irreflexive.

\item $\rightarrow^{irr}$; we define the {\em irreflexive horismos} in a similar way as we defined $\rightarrow$, this time without
permitting $x$ to be at horismos with itself.

\item $\ll^=$; we define the {\em reflexive chronology}  as we defined $\ll$, but this time we permit $x$ to
chronologically precede itself.

\item $\prec$; the {\em causal} order is a reflexive partial order defined as $x \prec y$ if $y \in C_+^T(x) \cup C_+^L(x)$.

\item $\ll^{\rightarrow^{irr}}$; we define the {\em irreflexive causal order} as we defined $\prec$, this time excluding
the case that $x \prec x$.

\item $\leq$; we define the {\em reflexive chorology} as we defined $<$, but this time we permit $x$ to
chorologically precede itself.

\item $\ll^c$; the {\em complement of chronological} order is a reflexive partial order defined as $x \ll^c y$ if $y \in C_+^S(x) \cup C_+^L(x)$.

\item $<^{\rightarrow^{irr}}$; we define the {\em irreflexive complement of chronological order} as $\ll^c$ excluding
the case that $x \ll^c x$.

\end{enumerate}





\begin{definition}
Let $f: M \to M$ be an one-to-one (and not necessarily continuous or linear) map.
We say
\begin{enumerate}

\item $f$ is a causal automorphism, if both
$f$ and $f^{-1}$ preserve $\ll$, i.e. $x \ll y$
iff $f(x) \ll f(y)$ and

\item $f$ is an acausal automorphism, if both
$f$ and $f^{-1}$ preserve $<$, i.e. $x < y$
iff $f(x) < f(y)$.

\end{enumerate}
\end{definition}

The causal automorphisms form the {\em causality group}
and the acausal automorphisms form the {\em acausality group}.

The proofs of lemmas \ref{1} and \ref{2} can be found
in \cite{Zeeman2}.

\begin{lemma}\label{1}

Let $f : M \to M$ be an one to one map. Then,
$f, f^{-1}$ preserve $\ll$ iff $f,f^{-1}$ preserve
$\rightarrow$. 
\end{lemma}

Lemma \ref{1} does not hold for $<$, for the obvious reason
that $x \rightarrow y$, iff either $x$ does not chronologically
precede $y$ or $y \ll z$ implies $x \ll z$. Consequently, Lemma \ref{1}
does not hold for the relations numbered 8, 9 and 10, above, while
it holds for the relations 2, 6 and 7.

\begin{lemma}\label{2}
A causal automorphism maps:
\begin{enumerate}
\item light rays to light rays;

\item parallel light rays to parallel light rays;

\item each light ray linearly and

\item parallel equal intervals on light rays to
parallel equal intervals.

\end{enumerate}
\end{lemma}

Lemma \ref{2} does not hold for an acausal automorphism,
for similar reasons that $<$ fails to satisfy Lemma \ref{1}.

The {\em orthochronous} Lorentz group consists of all linear
maps of $M$ which leave $Q$ invariant, preserve time orientation (South-to-North)
but possible reverse space orientation. In the $2$-dimensional
Minkowski space $M$, the {\em orthochorous} Lorentz
group consists of all linear maps of $M$ which leave $Q$ invariant,
preserve space-orientation (West-to-East) but possibly reverse time orientation.

\section{Thirty Causal Topologies on the $2$-dimensional Minkowski Space.}

Consider an order relation $R$ defined on a space $X$. Then, consider the sets $I^+(x) = \{y \in X : x R y\}$
and $I^-(x) = \{y \in X : y R x\}$, as well as the collections $\mathcal{S}^+ = \{X\setminus I^-(x) : x \in X\}$
and $\mathcal{S}^- = \{X\setminus I^+(x) : x \in X\}$. A basic-open set $U$ in the {\em interval topology} $T^{in}$
(see \cite{Compendium})
is defined as $U = A \cap B$, where $A \in \mathcal{S}^+$ and $B \in \mathcal{S}^-$; that is,
 $\mathcal{S}^+ \cup \mathcal{S}^-$ forms a subbase for $T^{in}$.

The $4$-dimensional Minkowski space in particular (and spacetimes in general)
is not up-complete, and a topology $\cT_{in}$ is weaker than the interval topology of \cite{Compendium},
but
for the particular case of $2$-dimensional Minkowski spacetime,
$\cT_{in}$ under the ten causal relations that we stated above
is the actual interval topology defined in \cite{Compendium}.

The {\em Alexandrov} topology (see \cite{Penrose-difftopology}) is the
topology which has basic open sets of the form $I^+(x) \cap I^-(y)$,
where $I^+(x) = \{y \in M : x \ll y\}$ and $I^-(y) = \{x \in M : y \ll x\}$.
In general, a spacetime manifold $M$ is strongly causal iff the Alexandrov topology
is Hausdorff iff the Alexandrov topology agrees with the manifold topology.

Last, but not least,
If $T_1$ and $T_2$ are two distinct topologies on a set $X$, then the {\em intersection topology} $T^{int}$ (see \cite{Intersection})
with respect to $T_1$ and $T_2$, is the topology on $X$ such that the set $\{U_1 \cap U_2 : U_1 \in T_1, U_2 \in T_2\}$
forms a base for $(X,T)$.

Below, we list all possible order topologies that are generated
by the ten causal relations above, either by defining the topology straight
from the order (in a similar way the Alexandrov topology is induced by $\ll$-open
diamonds) or as interval topologies $\cT_{in}$ or as intersection topologies (in the sense of Reed)
between the natural topology $E$ of $\mathbb{R}^2$ and $\cT_{in}$.

\begin{enumerate}

\item[1.] The chronological order $\ll$ induces the topology $T_{\ll}$,
which has a subbase consisting of future time cones $C_+^T(x)$ or past time cones $C_-^T(y)$,
where $x,y \in M$. The finite intersections of such subbasic-open sets give
``open timelike diamonds'', which are basic-open sets for the Alexandrov topology.

\item[2.] $\ll$ also induces the interval topology $T_{in}^{\ll}$,
with subbase consisting of sets $M\setminus C_+^T(x)$, which are complements of future time cones
or sets $M\setminus C_-^T(x)$ which are complements of past time cones. This topology
has basic-open sets of the form $C^S(x) \cup C^L(x)$ and it is
easy to see that it is incomparable (neither finer, nor coarser, nor equal) to the natural topology $E$, on $M$.

\item[3.] The topologies $E$ and $T_{in}^{\ll}$, on $M$, give the
intersection topology $Z_{in}^{\ll}$, which has basic-open sets
of the form $B_\epsilon(x) \cap [C^S(x) \cup C^L(x)]$ and is finer
than the topology $E$.

\item[4.] The relation horismos $\rightarrow$ induces the topology $T_{\rightarrow}$,
which has a subbase consisting of future light cones $C_+^L(x) \cup \{x\}$ or past light cones $C_-^L(y) \cup \{y\}$,
where $x,y \in M$. The finite intersections of such subbasic-open sets give the
boundaries of
``open diamonds'' that we examined in topology 1.

\item[5.] $\rightarrow$ also induces the interval topology $T_{in}^{\rightarrow}$,
with subbase consisting of sets $M\setminus [C_+^L(x) \cup \{x\}]$, which are complements of future light cones union $\{x\}$
or sets $M\setminus [C_-^L(x) \cup \{x\}]$ which are complements of past light cones union $\{x\}$. This topology
has basic-open sets of the form $[C^S(x) \cup C^T(x)] \setminus \{x\}$ and it is
incomparable to the natural topology of $M$.

\item[6.] The topologies $E$ and $T_{in}^{\rightarrow}$, on $M$, give the
intersection topology $Z_{in}^{\rightarrow}$, which has basic-open sets
of the form $B_\epsilon(x) \cap [(C^S(x) \cap C^T(x)) \setminus \{x\}]$
and is a finer topology than $E$.

\item[7.] The chorological order $<$ induces the topology $T_<$,
which has a subbase consisting of $+$-oriented (and deleted by definition, i.e. not including $x$) space cones $C_+^S(x)$ or $-$-oriented (deleted) space cones $C_-^S(y)$,
where $x,y \in M$. The finite intersections of such subbasic-open sets give
``open diamonds'' that are spacelike.

\item[8.] $<$ induces the interval topology $T_{in}^<$,
with subbase consisting of sets $M\setminus C_+^S(x) $, which are complements of $+$-oriented space cones
or sets $M\setminus C_-^S(x)$ which are complements of $-$-ve oriented  space cones. This topology
has basic-open sets of the form $C^T(x) \cup C^L(x)$ (causal cones) and it is
easy to see that it is incomparable to the natural topology of $M$.

\item[9.] The topologies $E$ and $T_{in}^<$, on $M$, give the
intersection topology $Z_{in}^<$, which has basic-open sets
of the form $B_\epsilon(x) \cap [C^T(x) \cup C^L(x)]$ and is a topology
finer than $E$.

\item[10.] The irreflexive horismos $\rightarrow^{irr}$ induces the topology $T_{\rightarrow^{irr}}$,
which has a subbase consisting of deleted (that is, without $\{x\}$ future light cones $C_+^L(x)\setminus \{x\}$ or deleted past light cones $C_-^S(y)\setminus \{y\}$,
where $x,y \in M$. The finite intersections of such subbasic-open sets give
deleted boundaries of ``open diamonds''.

\item[11.] $\rightarrow^{irr}$  induces the interval topology $T_{in}^{\rightarrow^{irr}}$,
with subbase consisting of sets $M\setminus [C_+^L(x) \setminus \{x\}]$, which are complements of deleted future light cones
or sets $M\setminus [C_-^L(x) \setminus \{x\}]$ which are complements of deleted past light cones. This topology
has basic-open sets of the form $[C^T(x) \cup C^S(x)] \cup \{x\}$ and it is
easy to see that it is incomparable to the natural topology of $M$.

\item[12.] The topologies $E$ and $T_{in}^{\rightarrow^{irr}}$, on $M$, give the
intersection topology $Z_{in}^{\rightarrow^{irr}}$, which has basic-open sets
of the form $B_\epsilon(x) \cap [(C^T(x) \cup C^S(x)) \cup \{x\}]$ and is a topology
finer than $E$.

\item[13.] The reflexive chronology $\ll^{=}$ induces the topology $T_{\ll^{=}}$,
which has a subbase consisting of future time cones $C_+^T(x) \cup \{x\}$ or past time cones $C_-^T(y) \cup \{y\}$,
where $x,y \in M$. The finite intersections of such subbasic-open sets give
``closed diamonds'', in the sense of a closed interval containing its endpoints.

\item[14.] $\ll^{=}$  induces the interval topology $T_{in}^{\ll^{=}}$,
with subbase consisting of sets $M\setminus [C_+^T(x) \cup \{x\}]$,
or sets $M \setminus [C_-^T(x) \cup \{x\}]$. This topology
has basic-open sets of the form $[C^S(x) \cup C^L(x)] \setminus \{x\}$ and it is incomparable to the natural topology of $M$.

\item[15.] The topologies $E$ and $T_{in}^{\ll^{=}}$, on $M$, give the
intersection topology $Z_{in}^{\ll^{=}}$, which has basic-open sets
of the form $B_\epsilon(x) \cap [(C^S(x) \cup C^L(x)) \setminus \{x\}]$ and it is
a topology finer than $E$.

\item[16.] The irreflexive causal order $\ll^{\rightarrow^{irr}}$ induces the topology $T_{\ll^{\rightarrow^{irr}}}$,
which has a subbase consisting of (deleted) future causal cones $[C_+^T(x) \cup C_+^L(x)] \setminus \{x\}$ or (deleted) past causal
 cones $[C_-^T(y) \cup C_-^L(y)] \setminus \{y\} $,
where $x,y \in M$. The finite intersections of such subbasic-open sets give
``causal diamonds'' which are open (causal diamonds, i.e. together with their light boundaries), but
without the endpoints.

\item[17.] $\ll^{\rightarrow^{irr}}$  induces the interval topology $T_{in}^{\ll^{\rightarrow^{irr}}}$,
with subbase consisting of sets $M\setminus [C_+^T(x) \cup C_+^L(x) \setminus \{x\}]$, which are complements of deleted future causal cones
or sets $M\setminus [C_-^T(x) \cup C_-^L(x) \setminus \{x\}]$ which are complements of deleted past causal cones. This topology
has basic-open sets of the form $C^S(x)$, that is space cones, and it is
easy to see that it is incomparable to the natural topology of $M$.

\item[18.] The topologies $E$ and $T_{in}^{\ll^{\rightarrow^{irr}}}$, on $M$, give the
intersection topology $Z_{in}^{\ll^{\rightarrow^{irr}}}$, which has basic-open sets
of the form $B_\epsilon(x) \cap C^S(x)$ (bounded space cones) and it is finer than
$E$.

\item[19.] The causal order $\prec$ induces the topology $T_{\prec}$,
which has a subbase consisting of future causal cones $C_+^T(x) \cup C_+^L(x)$ or past causal cones $C_-^T(y) \cup C_-^L(y)$,
where $x,y \in M$. The finite intersections of such subbasic-open sets give
``causal diamonds'', containing the endpoints.

\item[20.] $\prec$ induces the interval topology $T_{in}^{\prec}$,
with subbase consisting of sets $M\setminus [C_+^T(x) \cup C_+^L(x)]$, which are complements of future causal cones
or sets $M\setminus [C_-^T(x) \cup C_-^L(x)]$ which are complements of past causal cones. This topology
has basic-open sets of the form $C^S(x) \setminus \{x\}$ and it is
easy to see that it is incomparable to the natural topology of $M$.

\item[21.] The topologies $E$ and $T_{in}^{\prec}$, on $M$, give the
intersection topology $Z_{in}^{\prec}$, which has basic-open sets
of the form $B_\epsilon(x) \cap [C^S(x) \setminus \{x\}]$ and it is finer
than $E$.

\item[22.] The reflexive chorological order $\leq$ induces the topology $T_\leq$,
which has a subbase consisting of $+$-oriented space cones $C_+^S(x)$ or $-$-oriented space cones $C_-^S(y)$,
where $x,y \in M$. The finite intersections of such subbasic-open sets give
``closed diamonds'', that is diamonds containing the endpoints, that are spacelike.

\item[23.] $\leq$ induces the interval topology $T_{in}^\leq$,
with subbase consisting of sets $M\setminus [C_+^S(x) \cup \{x\}]$,
or sets $M\setminus [C_-^S(x) \cup \{x\}]$. This topology
has basic-open sets of the form $[C^T(x) \cup C^L(x)] \setminus \{x\}$ and it is
easy to see that it is incomparable to the natural topology of $M$.

\item[24.] The topologies $E$ and $T_{in}^\leq$, on $M$, give the
intersection topology $Z_{in}^\leq$, which has basic-open sets
of the form $B_\epsilon(x) \cap [(C^T(x) \cup C^L(x)) \setminus \{x\}]$ and it
is a finer topology than $E$.

\item[25.] The irreflexive complement of the chronological order, namely $<^{\rightarrow^{irr}}$, induces the topology $T_{<^{\rightarrow^{irr}}}$,
which has a subbase consisting of $+$-oriented (deleted) space cones with their light boundary
$[C_+^S(x) \cup C_+^L(x)] \setminus \{x\}$ or $-$-oriented (deleted) space cones with their light boundary $[C_-^S(y) \cup C_-L(y)] \setminus \{y\}$,
where $x,y \in M$. The finite intersections of such subbasic-open sets give
deleted ``open diamonds'' that are spacelike.

\item[26.] $<^{\rightarrow^{irr}}$ induces the interval topology $T_{in}^{<^{\rightarrow^{irr}}}$,
with subbase consisting of sets $M\setminus [C_+^S(x)\cup C_+^L(x) \setminus \{x\}]$, or sets $M\setminus [C_-^S(x) \cup C_-^L(x) \setminus \{x\}]$. This topology
has basic-open sets of the form $C^T(x)$, i.e. time cones, and it is
easy to see that it is incomparable to the natural topology of $M$.

\item[27.] The topologies $E$ and $T_{in}^{<^{\rightarrow^{irr}}}$, on $M$, give the
intersection topology $Z_{in}^{<^{\rightarrow^{irr}}}$, which has basic-open sets
of the form $B_\epsilon(x) \cap C^T(x)$. This intersection topology is the special relativistic analogue of the Path topology, introduced in \cite{Hawking-Topology} and it is finer than $E$.

\item[28.] The complement of the chronological order, namely ${\ll}^c$, induces the topology $T_{{\prec}^a}$,
which has a subbase consisting of $+$-oriented space cones with their light boundary
$C_+^S(x) \cup C_+^L(x)$ or $-$-oriented space cones with their light boundary $C_-^S(y) \cup C_-L(y)$,
where $x,y \in M$. The finite intersections of such subbasic-open sets give
 ``closed diamonds'' that are spacelike.

\item[29.] ${\ll}^c$ induces the interval topology $T_{in}^{{\ll}^c}$,
with subbase consisting of sets $M\setminus [C_+^S(x)\cup C_+^L(x)]$, which are complements of $+$-oriented space cones
with their light boundary or sets $M\setminus [C_-^S(x) \cup C_-^L(x)]$ which are complements of $-$-ve oriented space cones with their light boundary. This topology
has basic-open sets of the form $C^T(x) \setminus \{x\}$, i.e. deleted time cones, and it is
easy to see that it is incomparable to the natural topology of $M$.

\item[30.] The topologies $E$ and $T_{in}^{{\prec}^a}$, on $M$, give the
intersection topology $Z_{in}^{{\prec}^a}$, which has basic-open sets
of the form $B_\epsilon(x) \cap (C^T(x) \setminus \{x\})$ and it is finer than $E$.

\end{enumerate}

\section{Discussion.}

\subsection{Curved Spacetimes.}
A first question is if one can generalise
the thirty above mentioned topologies to curved spacetimes; the
answer is positive.
Indeed, from a topological perspective, and without any extra condition or
restriction one can
consider the general relativistic analogue of each
one of the mentioned topologies 1-30, since as soon as there
exists spacetime there are events and for each event there
is time/light/causal cone assigned to it; the point-set topology is
independent of the curvature and the tilt of the cones and since 
the mentioned topologies are generated from the causal relations
of the spacetime, one has to only choose an arbitrary Riemannian
metric $h$, on the spacetime manifold $M$. For example, the Path
topology of Hawking-King-McCarthy (see \cite{Hawking-Topology})
will be the generalisation of topology 3 of our list, as follows.

Consider the chronological order $\ll$, on a relativistic spacetime
manifold $M$. Then, $\ll$ will induce the interval topology $T_{in}^{\ll}$,
with subbase consisting of sets $M\setminus C_+^T(x)$, which are complements of future time cones
or sets $M\setminus C_-^T(x)$ which are complements of past time cones. This topology,
exactly as with our topology 3 of the list, 
has basic-open sets of the form $C^S(x) \cup C^L(x)$. Now, consider
the manifold topology $\mathcal{M}$ and for a Riemannian
metric $h$ consider the base of $\mathcal{M}$-open sets
of the form $B_\epsilon^h(x)$, the open balls centered at
$x$ and radius $\epsilon$ with respect to $h$. Then,
a basic-open set for the Path topology will be of the
form $T_{in}^{\ll} \cap B_\epsilon^h(x)$.
Low (see \cite{Low_path}) has shown that the Limit Curve
Theorem fails to hold for the Path topology, and so
the formation of a basic contradiction present in 
the proofs of all singularity theorems, fails as well
(for a more extensive discussion see \cite{Topology-Ambient-Boundary-Convergence},
\cite{Ordr-Ambient-Boundary} and \cite{Spacetimes-As-TS}). 

\subsection{Singularities.}
Furthermore, we observe that the Limit Curve Theorem holds for 
 each of the topologies $2,3, 8,9,14,15,23,24$ of our list, but
not for the topologies $5,6,11,12,17,18,20,21,26,27,29,30$. Following the
argument of Low (\cite{Low_path}, paragraph V), we can easily see if $U$ is a basic-open set
of either of the topologies $2,3, 8,9,14,15,23$ or $24$, then this
set does not contain the light cone of the event which defines it. Consider 
a sequence of null vectors $p_n$ converging to $p$ in the usual topology. Let
$\gamma_n$ be the null geodesic through the origin with tangent $p_n$ and
$\gamma$ the null geodesic through the origin with tangent vector $p$. Clearly,  
 $\gamma$ is the unique limit curve of the sequence $\{\gamma_n\}$
in the usual topology, for all $n$. But $\gamma_n$ intersected
with an open set (not containing the origin) of either of the basic-open sets defined in $2,3, 8,9,14,15,23$ or $24$ will give empty set,
so  $\gamma$ will be not a limit curve of the sequence $\gamma_n$ under the specified
topology either $2,3, 8,9,14,15,23$, or $24$ and so the Limit Curve Theorem will fail
for each of these topologies. On the contrary, following the same argument, the Limit Curve Theorem 
will hold for each of the topologies $5,6,11,12,17,18,20,21,26,27,29$ or $30$, since
each of them have basic-open sets containing the light-cone for each event.

\subsection{Ambient Cosmology.}

The significance of the above remarks is that one can construct topologies
which, unlike the manifold topology (which merely characterises continuity
properties according to Hawking et al.), there are thirty topologies (those
listed in this article) which determine the causal and conformal structures
of space-time and are most appealing than the Fine topology of Zeeman (which does
not admit a countable base of open sets). In addition, there are no other topologies
that can be defined immediately from the causal relations in a spacetime.

A question that is now raised is which topology is the most appropriate
one, if one can set it in this way, or the most physical one; the remark
that for eight of these topologies the Limit Curve Theorem fails to hold,
could bring the discussion on the need for an Ambient Cosmology to a
different level. For example, the very own construction of the ambient boundary-ambient
space model (see \cite{Cotsakis1}) was an attempt to consider a $4$-dimensional spacetime
as the conformal infinity of a $5$-dimensional ambient space, to show that singularities
are absent and the Cosmic Censorship becomes valid by construction. In
the frame of topologies like some ones that we mentioned in this paper
though, this is achieved without the need of working in extra dimensions (all due to the reason that LCT fails under them).

Last, but not least, the topologies from our list could be linked to the study of sliced spaces (see \cite{sliced}). In particular, a sliced spacetime $V$ can be considered as a product of a smooth manifold $M$ of dimension $3$ times the real line $\mathbb{R}$, where $V$ is equipped with with a $4$-dimensional Lorentzian metric that splits in a particular way with respect to a shift and a lapse function. So far global hyperbolicity has been studied with respect to the Alexandrov topology and the natural Produc topology of $V$, but not with respect to causal topologies. This would be particularly interesting in the case of Ambient Cosmology, since one can consider $M$ as a $4$-dimensional spacetime, the conformal boundary at infinity of an ambient space $V = M \times \mathbb{R}$.

\subsection{Girders and Twistor Spaces.}
Lastly, topologies $4, 10$ seem to fit well in spaces consisted of girders, hypergirders
and links (see \cite{Penrose-Kronheimer}). Although they depend on the structure of
the light cone, the question that has to be addressed is how they could be used in
a description of the transition from  quantum non-local theory to a classical local theory.
Certainly, there is not a definite answer to this question at the present moment but
we believe that methods of point-set topology will contribute significantly,
as one can work using topological tools invariantly from the geometry of a spacetime. Topologies that are constructed through light-rays, like $4$ and $10$ in our list, could also place a role in twistor theory (see \cite{Road-to-reality}). More particularly, two points $P,R$ in the Minkowski space that are incident with the same non-zero (null) twistor $Z$  must be null-separated by each other which means that $Z$ defines a light ray in the Minkowski space. This gives an inside of how one could tranfer topological properties from the Minkowski space to the twistor space $\mathbb{T}$ or, better, to the projective twistor space $\mathbb{PT}$.

\subsection{On Abstract Conformally Invariant Boundary Constructions in Relativity Theory.}

Some other possible applications of the topologies in the list of Section 3 might appear for abstract conformally invariant constructions in Relativity theory, namely boundaries of Low, Geroch, Kronheirmer and Penrose (see for example \cite{bound1}, \cite{bound2}, \cite{bound3}, \cite{bound4}, \cite{bound5} and \cite{bound6}). These topologies are based purely on the causal structure of the spacetime, but it is not clear which one would be the canonical choice of the topology. Low has proposed a new such topology in \cite{bound3} where the causal boundary has been developed further in relation to the classical conformal boundary. A further study could be based, for example, on the thorough study in \cite{bound4}, through links with deep constructions in pure mathematics in \cite{bound5} and on a recent revision in \cite{bound6}.

\section*{Acknowledgements.}
The authors are extremely grateful to a Reviewer who, apart from a careful proofreading and correcting the text from typos, he/she has opened to us new research directions; special thanks for the bibliography on abstract conformally invariant boundary constructions.

\end{document}